\documentclass{article}
\usepackage{ijcai26}

\usepackage{times}
\usepackage{soul}
\usepackage{url}
\usepackage[hidelinks]{hyperref}
\usepackage[utf8]{inputenc}
\usepackage[small]{caption}
\usepackage{graphicx}
\usepackage{amsmath}
\usepackage{amsthm}
\usepackage{booktabs}
\usepackage{algorithm}
\usepackage{algorithmic}
\usepackage[switch]{lineno}

\usepackage{cite}
\usepackage{tabularx}
\usepackage{subcaption}
\usepackage{multirow}
\theoremstyle{definition}
\newtheorem{dfn}{Definition}

\title{DPSQL+: A Differentially Private SQL Library with a Minimum Frequency Rule}

\author{
Tomoya Matsumoto$^1$\thanks{These authors contributed equally to this work.}
\and
Shokichi Takakura$^1$\footnotemark[1]
\and
Shun Takagi$^1$
\and
Satoshi Hasegawa$^1$
\\
\affiliations
$^1$LY Corporation\\
\emails
\{tomoymat, stakakur, shutakag, satoshi.hasegawa\}@lycorp.co.jp
}

\begin{document}

\maketitle

\begin{abstract}
SQL is the de facto interface for exploratory data analysis; however, releasing exact query results can expose sensitive information through membership or attribute inference attacks.
Differential privacy (DP) provides rigorous privacy guarantees, but in practice, DP alone may not satisfy governance requirements such as the \emph{minimum frequency rule}, which requires each released group (cell) to include contributions from at least $k$ distinct individuals.
In this paper, we present \textbf{DPSQL+}, a privacy-preserving SQL library that simultaneously enforces user-level $(\varepsilon,\delta)$-DP and the minimum frequency rule.
DPSQL+ adopts a modular architecture consisting of: (i) a \emph{Validator} that statically restricts queries to a DP-safe subset of SQL; (ii) an \emph{Accountant} that consistently tracks cumulative privacy loss across multiple queries; and (iii) a \emph{Backend} that interfaces with various database engines, ensuring portability and extensibility.
Experiments on the TPC-H benchmark demonstrate that DPSQL+ achieves practical accuracy across a wide range of analytical workloads---from basic aggregates to quadratic statistics and join operations---and allows substantially more queries under a fixed global privacy budget than prior libraries in our evaluation.
\end{abstract}

\section{Introduction}
\label{sec:introduction}

While SQL remains the de facto interface for exploratory analytics, releasing exact query answers poses significant privacy risks, including membership and attribute inference~\cite{Dinur2003Revealing,Dwork2010Difficulties}. 
Differential privacy (DP) provides a rigorous privacy guarantee~\cite{Dwork2014Algorithmic}, yet integrating DP into practical SQL workflows remains challenging in real-world deployments.

Existing DP-SQL systems~\cite{Johnson2020Chorus,Google2023ZetaSQL,OpenDP2023SmartNoise,ByteDance2023Jeddak,Grislain2024Qrlew,Yu2024DOP-SQL} exhibit several important limitations. 
First, DP alone does not enforce the ``minimum frequency rule'' (threshold rule)~\cite{Sukasih2012Implementing,Garfinkel2023De-Identifying} required by many data governance frameworks, meaning that DP-compliant outputs may still violate institutional suppression policies. 
Second, some systems do not implement privacy accounting~\cite{Dwork2014Algorithmic} as a built-in component, making consistent and compositional privacy budget management difficult. 
Third, extensibility across heterogeneous SQL backends is limited, making the integration of new backends difficult. 
Moreover, support is restricted to a narrow class of queries, preventing the handling of common analytical constructs such as JOIN operations and thereby limiting applicability to real-world SQL workloads. 
Finally, for quadratic statistics such as variance and covariance, noise effects tend to be amplified, often resulting in substantial estimation error~\cite{Takakura2025Optimal}.

In this paper, we present \textbf{DPSQL+}\footnote{Code available at \url{https://github.com/lycorp-jp/DPSQL_Plus}.}, a library designed to address these limitations. 
DPSQL+ employs a modular architecture that separates (1) \emph{validation} of DP-safe SQL subsets, (2) \emph{integrated accounting} for multi-query sessions, and (3) \emph{backend execution} for noisy aggregation. 

\paragraph{Contributions.}
Our primary contributions are as follows:
\begin{itemize}
  \item \textbf{Unified Enforcement of DP and the Minimum Frequency Rule:} DPSQL+ combines $(\varepsilon, \delta)$-DP with a threshold constraint, ensuring that every released group contains at least $k$ distinct users, thereby satisfying both formal DP and regulatory requirements.

  \item \textbf{Embedded Multi-Query Privacy Accounting:} DPSQL+ incorporates privacy accounting directly into the system architecture, enabling global budget enforcement across multiple queries. It supports advanced composition techniques such as R\'enyi DP and Privacy Loss Distributions (PLD)~\cite{Bun2016Concentrated,Google2025PLD} for tighter composition.

  \item \textbf{Extensible Architecture with Broad SQL and Backend Support:} DPSQL+ uses a decoupled validation and execution framework that simplifies adding new SQL backends and enhances portability. It supports a wide range of queries, including JOIN operations, enabling practical use in real-world SQL workloads.

  \item \textbf{Optimized Mechanisms for Quadratic Statistics:} DPSQL+ implements optimized estimators for variance and covariance~\cite{Takakura2025Optimal}, improving practical analytical utility.
\end{itemize}

\section{Related Work}
\label{sec:related_work}

\begin{table*}[t]
\centering
\small
\setlength{\tabcolsep}{6pt}
\begin{tabularx}{\textwidth}{l c c c X}
\toprule
\textbf{System} &
\textbf{Architecture} &
\textbf{Min freq.} &
\textbf{Accountant} &
\textbf{Backend compatibility} \\
\midrule
\textbf{DPSQL+ (ours)} &
Proxy-based &
Yes &
Integrated &
Spark SQL, SQLite, DuckDB \\
\addlinespace

SmartNoise SQL~\cite{OpenDP2023SmartNoise} &
Proxy-based &
No &
Integrated &
Spark SQL, SQLite, PostgreSQL, BigQuery, SQL Server, etc \\
\addlinespace

Qrlew~\cite{Grislain2024Qrlew} &
Query Rewriting &
No &
External &
PostgreSQL, BigQuery, Synapse SQL \\
\addlinespace

ZetaSQL-DP~\cite{Google2023ZetaSQL} &
Engine Integration &
No &
External &
BigQuery \\
\addlinespace

DOP-SQL~\cite{Yu2024DOP-SQL} &
Engine Integration &
No &
External &
PostgreSQL \\
\addlinespace

Chorus~\cite{Johnson2020Chorus} &
Query Rewriting &
No &
Integrated &
PostgreSQL \\
\addlinespace

Jeddak-DPSQL~\cite{ByteDance2023Jeddak} &
Proxy-based &
No &
Integrated &
ClickHouse, Hive \\
\bottomrule
\end{tabularx}
\caption{Comparison of differentially private SQL systems. ``Accountant'' indicates whether privacy budget management across multiple queries is integrated as a system component.}
\label{tab:relatedwork-comparison}
\end{table*}

Existing DP-SQL systems (Table~\ref{tab:relatedwork-comparison}) fall into three architectural paradigms: query rewriting, engine integration, and proxy-based approaches.

\paragraph{Query Rewriting.}
Both \textbf{Chorus}~\cite{Johnson2020Chorus} and \textbf{Qrlew}~\cite{Grislain2024Qrlew} utilize a query-rewriting architecture, which transforms input SQL into DP-compliant forms. 
By operating independently of the database engine, this design enables seamless integration with existing analytical workflows. 
However, because it lacks access to internal execution semantics, the sensitivity analysis tends to be conservative. 
Furthermore, as noise generation is delegated to the underlying database engine, Qrlew remains vulnerable to floating-point precision attacks~\cite{Mironov2012Significance,Haney2022Precision}.

\paragraph{Engine Integration.}
Both \textbf{ZetaSQL-DP}~\cite{Google2023ZetaSQL} and \textbf{DOP-SQL}~\cite{Yu2024DOP-SQL} adopt an engine-integrated architecture.
By embedding DP mechanisms directly into a specific SQL engine, they enable tight optimization and efficient execution.
However, this tight coupling makes it difficult to support additional SQL backends, limiting extensibility.

\paragraph{Proxy-based.}
Proxy-based systems introduce an intermediate layer between clients and databases, improving multi-backend extensibility.
Existing libraries such as \textbf{SmartNoise SQL}~\cite{OpenDP2023SmartNoise} and \textbf{Jeddak-DPSQL}~\cite{ByteDance2023Jeddak} delegate aggregation processing to the underlying database engine to maintain performance. 
However, this approach significantly restricts the supported SQL grammar, limiting complex queries.

\paragraph{Summary of Gaps.}
In summary, existing DP-SQL solutions are constrained by conservative sensitivity analysis, tight coupling to specific backends, or limited query expressiveness. 
Our \textbf{DPSQL+} addresses these gaps through a modular proxy architecture. 
By decoupling aggregation processing from the underlying database engine, DPSQL+ supports a broad class of queries, while preserving both extensibility across backends and analytical utility. 

\section{System Overview}
\label{sec:system_overview}

\begin{figure}[t]
    \centering
    \includegraphics[width=\linewidth]{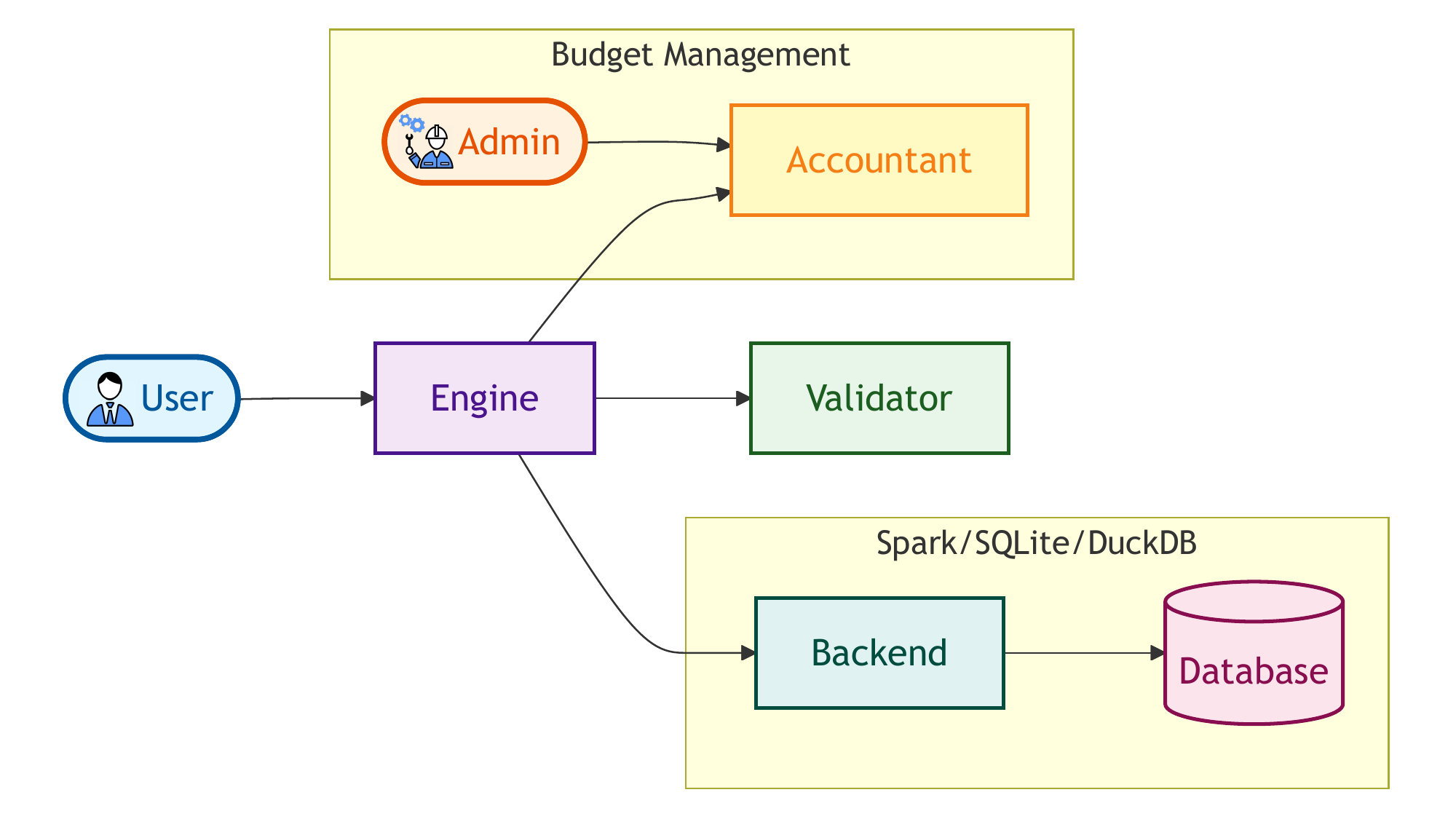}
    \caption{The architecture of DPSQL+.}
    \label{fig:architecture}
\end{figure}

DPSQL+ is a modular, proxy-based system designed to enforce user-level $(\varepsilon, \delta)$-DP and the minimum frequency rule for SQL-based analytics. 
The architecture, illustrated in Figure~\ref{fig:architecture}, mediates all interactions between analysts and underlying data stores through four primary components.

The \textbf{Engine} acts as the central orchestrator, managing the end-to-end workflow.
The \textbf{Validator} performs static analysis to parse SQL syntax and ensure adherence to privacy constraints, while the \textbf{Accountant} tracks cumulative privacy loss across queries.
Finally, the \textbf{Backend} interfaces with data engines such as Spark SQL or DuckDB to apply contribution bounding, double thresholding, and noisy aggregation.

When a query is submitted, the system guides it through four stages: validation, budget checking, execution, and delivery.
This automated pipeline ensures that all privacy rules are strictly followed and the global budget is updated before any results are released.
Ultimately, this removes the need for analysts to track privacy budgets manually, allowing them to focus entirely on their data analysis.

\section{Privacy Mechanisms}
\label{sec:privacy_mechanisms}

DPSQL+ implements mechanisms that satisfy both user-level $(\varepsilon,\delta)$-DP and the minimum frequency rule.
To achieve this, we extend the bounded user contribution framework~\cite{Wilson2020Differentially} with advanced mechanisms for thresholding and aggregation.
In this section, we formally define the privacy guarantees and detail the algorithms used for contribution bounding, double thresholding, noisy aggregation, and privacy accounting.

\subsection{Privacy Definitions}
We consider a dataset $D$ in which each record is associated with a user identifier, and each user may contribute multiple records. 
DPSQL+ guarantees user-level $(\varepsilon,\delta)$-DP under the add/remove adjacency relation (i.e., two datasets are adjacent if one can be obtained from the other by adding or removing all records of a single user).

\begin{dfn}[Differential privacy~\cite{Dwork2014Algorithmic}]
A randomized mechanism $\mathcal{M}$ satisfies $(\varepsilon, \delta)$-differential privacy if for any two adjacent datasets $D$, $D'$ and for all $S \subseteq \text{Range}(\mathcal{M})$:
\begin{equation}
\Pr[\mathcal{M}(D) \in S] \leq e^{\varepsilon} \cdot \Pr[\mathcal{M}(D') \in S] + \delta
\end{equation}
\end{dfn}

In addition to DP, strict regulations often require that no group is released unless it contains a sufficient number of distinct individuals.
DPSQL+ enforces this via a minimum frequency rule:
\begin{dfn}[Minimum frequency rule (Threshold rule)~\cite{Sukasih2012Implementing,Garfinkel2023De-Identifying}]
Let $U$ be the set of users and let $k$ be a positive integer threshold.
For each partition (group) $G$, define the set of distinct contributing users as
\[
C(G) := \{\, u \in U : u \text{ contributes to } G \,\}.
\]
The minimum frequency rule permits releasing the cell corresponding to $G$ only if
\[
|C(G)| \ge k.
\]
Otherwise, the cell is suppressed (not released).
\end{dfn}

\subsection{Core Algorithms}
The query execution follows a sequence of contribution bounding, double thresholding, and noisy aggregation.

\paragraph{Contribution Bounding.}
To bound the sensitivity of the query, DPSQL+ applies contribution bounding~\cite{Wilson2020Differentially}, which restricts the maximum number of rows a single user can contribute to the aggregation.
While prior works typically apply a one-time bound, DPSQL+ adopts the two-stage contribution bounding strategy proposed in Plume~\cite{Amin2022Plume}.
This method applies bounding both before and after the thresholding step.
As a result, it prevents users from consuming their limited contribution budget on keys that are eventually pruned (not output), thereby improving the utility of the released statistics.

\paragraph{Double Thresholding.}
A critical challenge in private SQL is determining which groups (keys) can be safely released.
Releasing all keys, including low-count or empty groups, can violate privacy.
Standard (single) $\tau$-thresholding~\cite{Wilson2020Differentially} releases groups only if their noisy count exceeds a threshold $\tau$; however, this approach does not guarantee that the \textit{true} count is at least $\tau$.
To satisfy the dual requirements of differential privacy and minimum frequency rule, DPSQL+ implements a two-stage thresholding logic derived from recent advancements in Gaussian sparse histogram mechanism~\cite{Wilkins2024Exact}.
This technique applies thresholding checks on both the exact (noiseless) count and the noisy count.
DPSQL+ uses $k$ as the threshold for the exact count, and $\tau$ as the threshold for the noisy count.
This ensures that the system enforces a minimum frequency constraint on the output while maintaining strict $(\varepsilon, \delta)$-DP.

\paragraph{Noisy Aggregation.}
Following bounding and thresholding, the data is aggregated, and noise is injected into the results.
While earlier systems~\cite{Wilson2020Differentially} utilized the Laplace mechanism, DPSQL+ employs the Gaussian mechanism.
This choice is motivated by favorable composition properties and compatibility with modern accounting frameworks.

\subsection{Advanced Aggregation Mechanisms}
Under the add-remove definition of adjacency, standard mechanisms for variance and covariance estimation often suffer from high error and instability when implemented via simple noisy sums of squares.
To improve utility, DPSQL+ integrates the B\'ezier mechanism~\cite{Takakura2025Optimal} for quadratic statistics (\texttt{VAR}, \texttt{STDDEV}, \texttt{COVAR}).
The B\'ezier mechanism constructs an optimal estimator that minimizes the mean squared error (MSE) for bounded range inputs, providing higher accuracy than standard approaches under the same privacy budget.

\subsection{Privacy Accounting}
DPSQL+ integrates a stateful accountant to track cumulative privacy loss, avoiding the loose bounds of basic sequential composition. 
The system supports two methods: R\'enyi differential privacy (RDP)~\cite{Bun2016Concentrated}, which is optimized for Gaussian mechanisms, and the Privacy Loss Distribution (PLD)~\cite{Google2025PLD}, which provides tighter bounds for heterogeneous workloads. 
This integrated approach automates budget management and maximizes the number of queries permissible within a fixed $(\varepsilon, \delta)$ limit, significantly enhancing practical utility.

\subsection{Floating-Point Safety}
Theoretical DP guarantees assume infinite-precision real numbers.
However, standard floating-point implementations can be vulnerable to attacks that exploit rounding behavior~\cite{Mironov2012Significance, Haney2022Precision}.
To mitigate this, DPSQL+ delegates low-level noise generation and aggregation logic to the OpenDP library~\cite{OpenDPLibrary}, which implements secure noise generation.

\begin{figure*}[!t]
  \centering
  \begin{subfigure}[t]{0.32\textwidth}
    \centering
    \includegraphics[width=\linewidth]{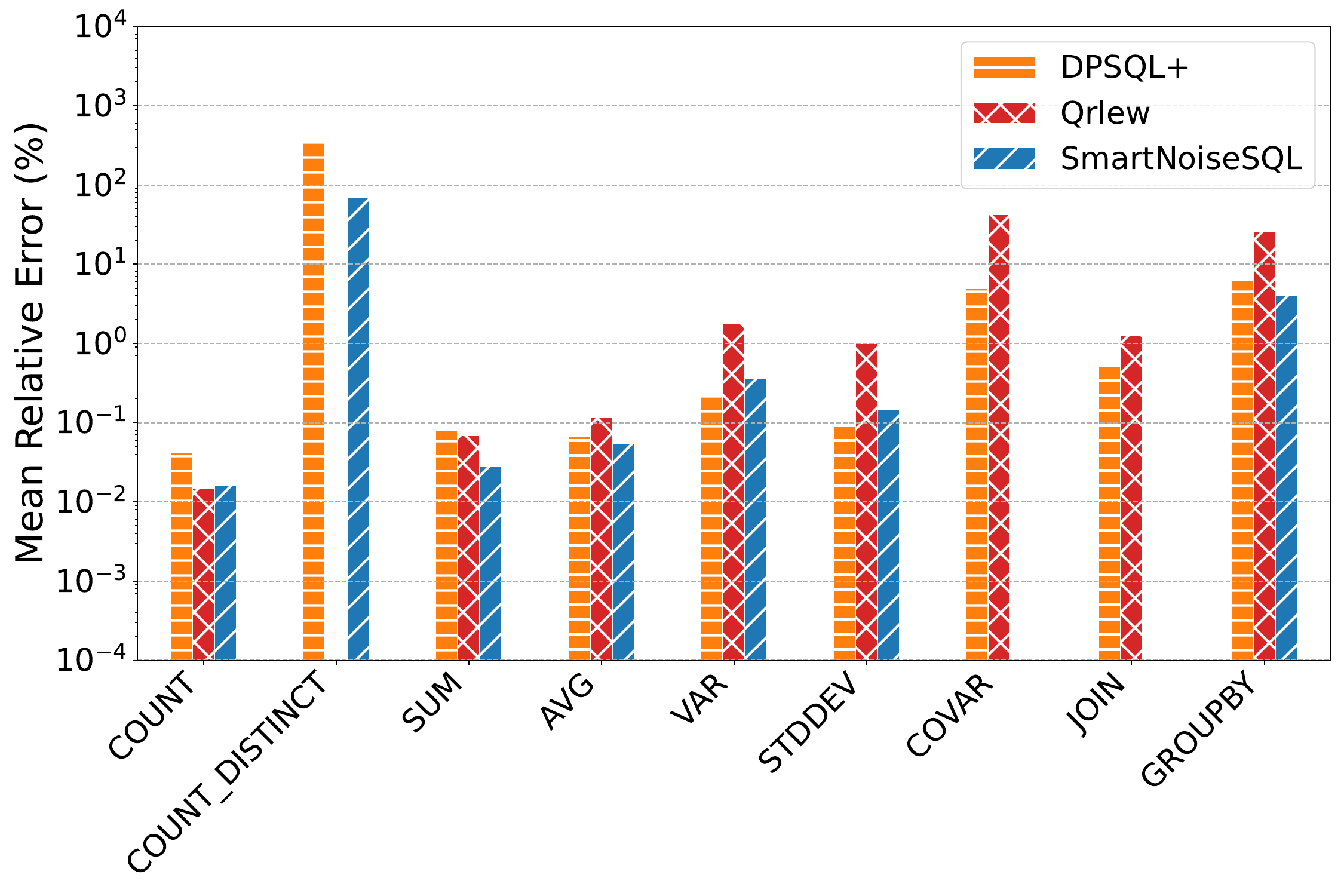}
    \caption{$\varepsilon = 0.1$}
  \end{subfigure}
  \hfill
  \begin{subfigure}[t]{0.32\textwidth}
    \centering
    \includegraphics[width=\linewidth]{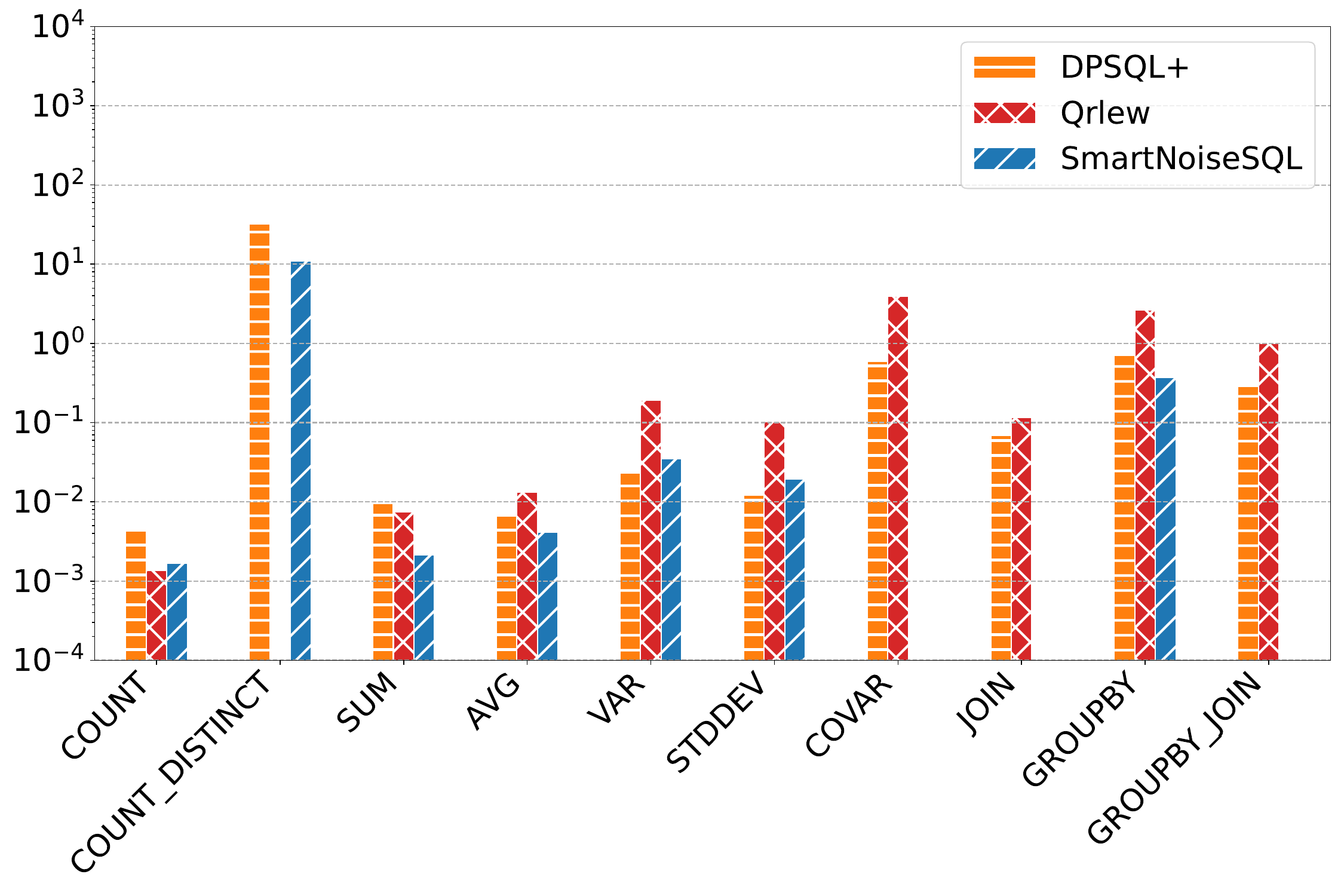}
    \caption{$\varepsilon = 1$}
  \end{subfigure}
  \hfill
  \begin{subfigure}[t]{0.32\textwidth}
    \centering
    \includegraphics[width=\linewidth]{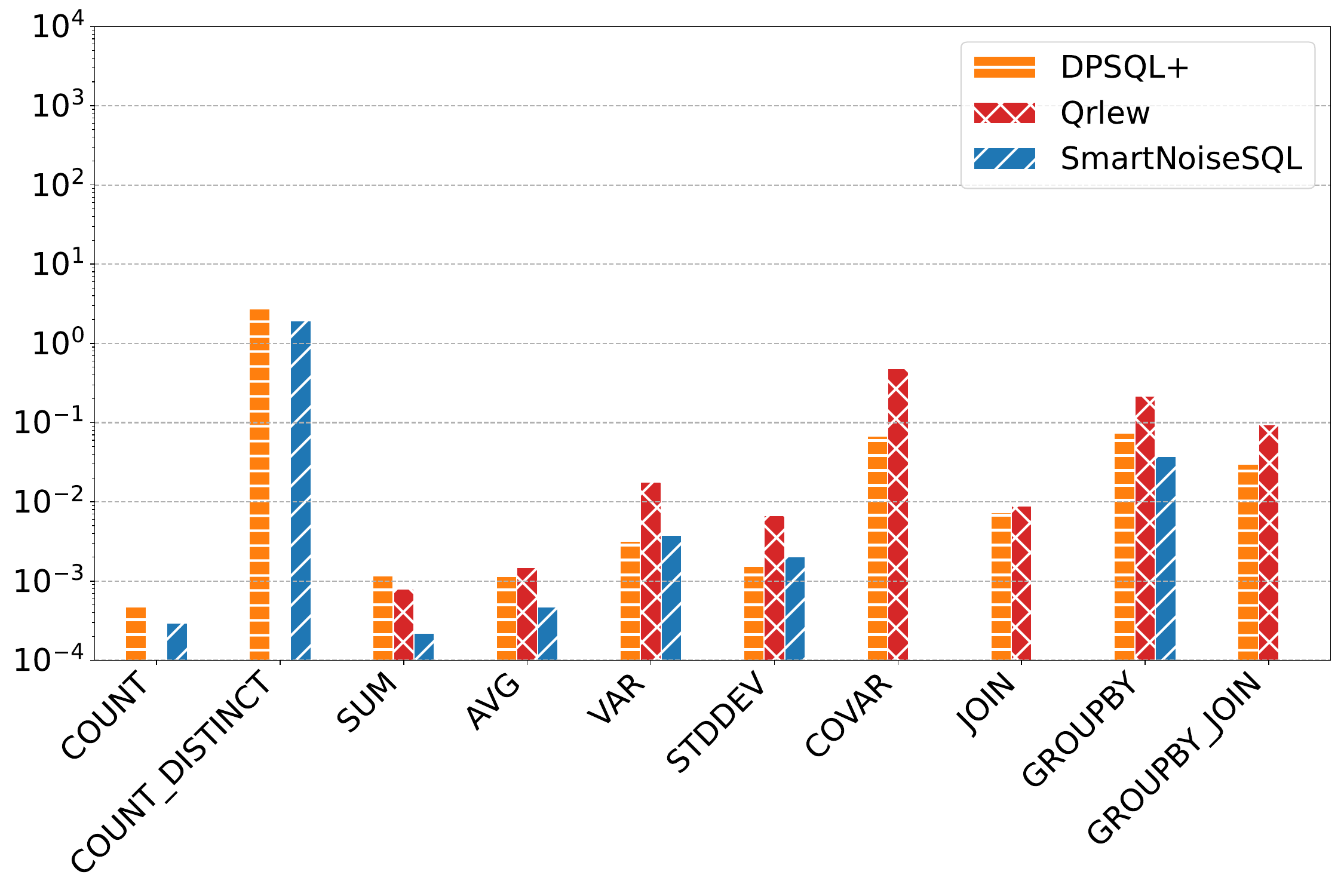}
    \caption{$\varepsilon = 10$}
  \end{subfigure}
  \caption{
  Mean Relative Error (\%) against the ground truth as a function of the privacy parameter $(\varepsilon, 10^{-7})$. 
  The y-axis is shown on a logarithmic scale.
  Qrlew does not support \texttt{COUNT\_DISTINCT}, and SmartNoise SQL does not support \texttt{COVAR}, \texttt{JOIN} and \texttt{GROUPBY\_JOIN}. 
  For $\varepsilon = 0.1$, \texttt{GROUPBY\_JOIN} is excluded because $\tau$-thresholding removes keys.
  }
  \label{fig:error_comparison}
\end{figure*}

\begin{figure}[!t]
  \centering
  \begin{subfigure}[t]{0.49\columnwidth}
    \centering
    \includegraphics[width=\linewidth]{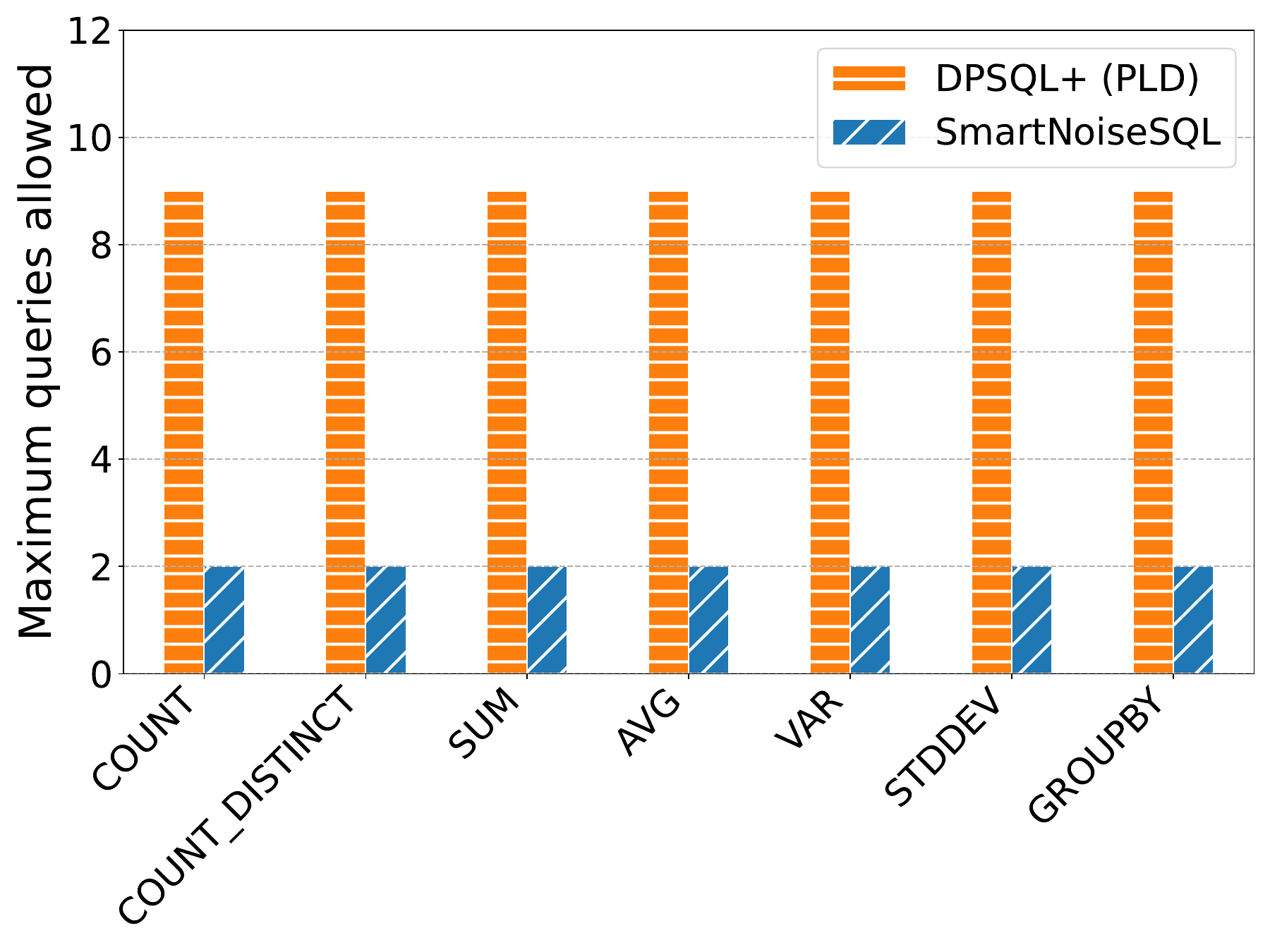}
    \caption{Global $(0.2, 10^{-6})$}
  \end{subfigure}
  \hfill
  \begin{subfigure}[t]{0.49\columnwidth}
    \centering
    \includegraphics[width=\linewidth]{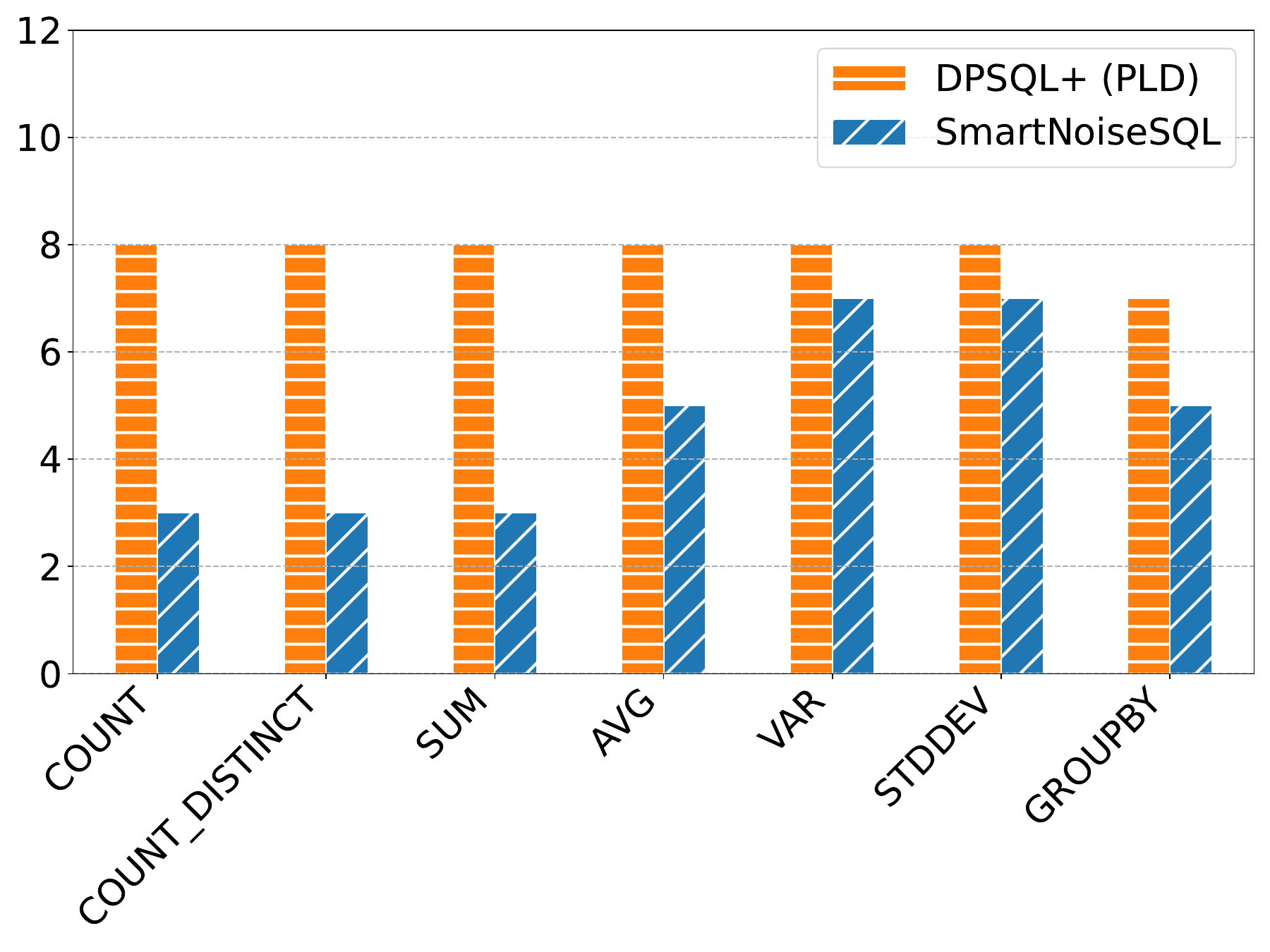}
    \caption{Global $(1, 2\times10^{-7})$}
  \end{subfigure}
  \caption{Maximum number of queries executable with fixed per-query budget $(\varepsilon, \delta)=(0.1, 10^{-7})$ under two global budgets.}
  \label{fig:accountant_comparison}
\end{figure}

\section{Query Syntax}
\label{sec:query_syntax}

DPSQL+ enforces privacy through structural validation of (a largely) standard SQL dialect.

\paragraph{Privacy Unit Specification.}
To guarantee user-level DP, schemas must define a \textit{privacy unit column} (e.g., \texttt{user\_id}), similar to ZetaSQL-DP~\cite{Google2023ZetaSQL} and AWS Clean Rooms~\cite{AWSCleanRooms}. 
This identifier is essential for performing contribution bounding and sensitivity analysis.

\paragraph{Structural Constraints.}
To support the mechanisms in Section~\ref{sec:privacy_mechanisms}, the validator enforces two primary constraints:

\begin{enumerate}
    \item \textbf{Aggregation-Only Output:} The outermost \texttt{SELECT} must consist solely of aggregate expressions (e.g., \texttt{COUNT}, \texttt{SUM}, \texttt{AVG}, \texttt{VAR}) to prevent individual record leakage.
    \item \textbf{CTEs over Subqueries:} To ensure robust static analysis, nested subqueries in the \texttt{FROM} clause are prohibited. Complex logic must be structured using Common Table Expressions (CTEs) via the \texttt{WITH} clause.
\end{enumerate}

By maintaining high compatibility with standard SQL, DPSQL+ lowers the adoption barrier, requiring only minimal modifications to existing analytical queries.

\section{Evaluation}
\label{sec:evaluation}

We evaluate the utility of DPSQL+ by comparing it with two widely used baselines, Qrlew~\cite{Grislain2024Qrlew} and SmartNoise SQL~\cite{OpenDP2023SmartNoise}, following the benchmarking methodology of \cite{Ecoffet2025Experiments}.

\subsection{Experimental Setup}

Experiments were conducted on the TPC-H database~\cite{TPC-H} (SF=1.0) using 10 query patterns (see Appendix for details). 
We designated \texttt{CUSTKEY} as the privacy unit and aligned contribution bounds and clipping thresholds across all systems to ensure a fair comparison. 
For DPSQL+, the minimum frequency was fixed to 1; this effectively imposes no constraint, consistent with the behavior of the baseline libraries. 

\subsection{Experimental Results}

Figure~\ref{fig:error_comparison} presents the Mean Relative Error (MRE, \%) across privacy levels, averaged over 25 independent trials. 
Note that the systems use different noise mechanisms: SmartNoise SQL uses the Laplace mechanism, whereas DPSQL+ and Qrlew are based on the Gaussian mechanism.

For single-query utility, DPSQL+ matches Qrlew on basic aggregates (\texttt{COUNT}, \texttt{SUM}, and \texttt{AVG}), while SmartNoise SQL is slightly more accurate. 
This is likely because the Laplace mechanism often yields better accuracy for single-shot queries. 
\texttt{COUNT\_DISTINCT} exhibits larger errors for all methods, mainly because small ground-truth values inflate relative error.

For higher-order statistics (\texttt{VAR}, \texttt{STDDEV}, and \texttt{COVAR}), DPSQL+ consistently attains the lowest error, attributable to improved formulations enabled by the B\'ezier mechanism. 
DPSQL+ also outperforms Qrlew on relational workloads (\texttt{JOIN} and \texttt{GROUPBY\_JOIN}), whereas SmartNoise SQL does not support these join-based queries.

Figure~\ref{fig:accountant_comparison} shows the maximum number of queries executable under a given global privacy budget. 
Here, DPSQL+ allows more queries than SmartNoise SQL in both the $\varepsilon$- and $\delta$-limited regimes, reflecting the more favorable composition bounds under $(\varepsilon,\delta)$-DP accounting.

\section{Limitations}
\label{sec:limitations}

While DPSQL+ provides robust private analytics, it entails operational and security trade-offs.

\paragraph{Operational Trade-offs.}
To ensure user-level $(\varepsilon,\delta)$-DP, DPSQL+ restricts query expressiveness (e.g., requiring aggregations and prohibiting nested subqueries) and introduces computational overhead via sensitivity analysis and contribution bounding. 
These constraints are often necessary to achieve rigorous privacy and frequency compliance.

\paragraph{Session-Scoped Accounting.}
The integrated accountant tracks privacy loss only for the lifetime of an engine instance. 
Since budget tracking is not persisted across system restarts, host applications must implement external state management to enforce long-term privacy budgets.

\paragraph{Scope of Privacy Guarantee.}
Similar to ZetaSQL-DP~\cite{Google2023ZetaSQL}, DPSQL+ focuses on algorithmic privacy against output-based inference. 
It does not mitigate physical side-channel attacks (e.g., timing analysis) or runtime-induced leakages (e.g., integer overflows), which require complementary security measures.

\section{Conclusion}
\label{sec:conclusion}

We presented DPSQL+, a modular SQL library that provides user-level $(\varepsilon,\delta)$-DP while enforcing a minimum frequency rule.
DPSQL+ combines double thresholding and the Gaussian mechanism with integrated RDP/PLD accounting.
Experiments on TPC-H show that DPSQL+ achieves practical accuracy across analytical workloads---from basic aggregates to higher-order statistics and join operations---and supports substantially more queries under a fixed global privacy budget than existing libraries.

Future work includes relaxing query constraints to improve expressiveness and developing automated privacy calibration mechanisms~\cite{Cheu2025Toward} to reduce analysts' operational burden.

\bibliographystyle{unsrt} 
\bibliography{main}

\appendix
\section{Details of Evaluation Settings}

\begin{table*}[t]
\centering
\caption{Specification of the query set used for the experiments.}
\label{tab:query_set}

\begin{tabularx}{\textwidth}{llX}
\toprule
\textbf{Type} & \textbf{Query} & \textbf{Description} \\
\midrule

\multirow{8}{*}{Scalar}
 & \texttt{COUNT} & The total number of tuples in the customer table. \\
 & \texttt{COUNT\_DISTINCT} & The number of distinct nation keys in the customer table. \\
 & \texttt{SUM} & The sum of account balances in the customer table. \\
 & \texttt{AVG} & The mean account balance in the customer table. \\
 & \texttt{VAR} & The variance of account balances in the customer table. \\
 & \texttt{STDDEV} & The standard deviation of account balances in the customer table. \\
 & \texttt{COVAR} & The covariance between price and discount in the lineitem table. \\
 & \texttt{JOIN} & The average order price for joined orders and customer tables under a selection predicate. \\

\midrule

\multirow{2}{*}{Grouped}
 & \texttt{GROUPBY} & The average order price grouped by order status in the orders table. \\
 & \texttt{GROUPBY\_JOIN} & The average order price grouped by segment for joined orders and customer tables.\\

\bottomrule
\end{tabularx}

\end{table*}

The experiments in Section~\ref{sec:evaluation} employ 10 query patterns, grouped into scalar statistical queries and grouped statistical queries, as summarized in Table~\ref{tab:query_set}.

Qrlew does not support \texttt{COVAR} and fails to execute \texttt{VAR} and \texttt{STDDEV} due to runtime errors. 
We therefore derive \texttt{COVAR}, \texttt{VAR}, and \texttt{STDDEV} from intermediate quantities obtained via \texttt{AVG}-based queries.
SmartNoise SQL does not support JOIN operations, which prevents it from executing relational queries where the privacy unit column (e.g., \texttt{CUSTKEY}) is not directly present in the target table, such as the \texttt{COVAR} query on the lineitem table.

To enable a meaningful evaluation of covariance estimation (\texttt{COVAR}), we preprocess the dataset by reordering the \texttt{L\_DISCOUNT} column in the lineitem table, as the original TPC-H data exhibits negligible correlation between \texttt{L\_DISCOUNT} and \texttt{L\_EXTENDEDPRICE}. 
Specifically, we reorder \texttt{L\_DISCOUNT} so that its Pearson correlation with \texttt{L\_EXTENDEDPRICE} is approximately $0.3$.

To ensure a fair comparison that isolates the privacy mechanisms themselves, we fix the contribution bounds and clipping thresholds for all systems using the true minimum and maximum values computed from the TPC-H database. 
Although production deployments typically require these bounds to be determined via private estimation~\cite{Cheu2025Toward}, this controlled setting ensures that differences in the experimental results are driven by noise addition and privacy accounting, rather than by parameter tuning.

Utility is quantified using the Mean Relative Error (MRE, \%). 
To account for the probabilistic nature of the sanitization process, each query $Q$ is executed over $N=25$ independent trials to ensure statistical reliability.
Let $T_{Q,j}$ denote the ground-truth (original) result for the $j$-th output component of query $Q$, and let $S_{Q,j}^{i}$ denote the corresponding sanitized result obtained in the $i$-th trial.
Here, $i \in \{1,\dots,N\}$ indexes trials, and $j \in \{1,\dots,n\}$ indexes output components (i.e., groups for \texttt{GROUP BY} queries).
We define the Mean Relative Error (MRE) as~\cite{Ecoffet2025Experiments}:
\begin{equation}
MRE(Q) = \frac{100}{N \times n} \sum_{i=1}^{N} \sum_{j=1}^{n} \frac{\left|S_{Q,j}^{i} - T_{Q,j}\right|}{T_{Q,j}}.
\end{equation}
For scalar queries that return a single aggregate value (e.g., \texttt{COUNT} or \texttt{SUM}), we have $n=1$, and the definition reduces to:
\begin{equation}
MRE_{\text{scalar}}(Q) = \frac{100}{N} \sum_{i=1}^{N} \frac{\left|S_{Q,1}^{i} - T_{Q,1}\right|}{T_{Q,1}}.
\end{equation}

\end{document}